\documentclass[%
 reprint,doublecolumn, notitlepage,
 amsmath,amssymb,
 aps
]{revtex4-1}

\usepackage{graphicx}% Include figure files
\usepackage{dcolumn}% Align table columns on decimal point
\usepackage{bm}% bold math
\usepackage{siunitx}%allows us to use math mode without italics
\usepackage{comment}% Allows blocking out large sections
\usepackage{multirow}

\begin{document}

\title[Cryogenic Thermal Modeling of Microwave High Density Signaling]{Cryogenic Thermal Modeling of Microwave High Density Signaling}

\author{Naomi Raicu$^1$, Tom Hogan$^2$, Xian Wu$^3$, Mehrnoosh Vahidpour$^3$, David Snow$^3$, Matthew Hollister$^4$, Mark Field$^3$}

\affiliation{
1. University of Wisconsin-Milwaukee, P.O. Box 413, Milwaukee, WI 53201 \\
2. Quantum Design, Inc., 10307 Pacific Center Court, San Diego, CA 92121\\
3. Rigetti Computing, 775 Heinz Avenue, Berkeley, CA 94710 \\
4. Fermi National Accelerator Laboratory, PO Box 500, Batavia, IL 60510-5011
}

\date{February 3, 2025}

\begin{abstract}

% arxiv abstract reduced to fit into arxiv character limit on abstract
Superconducting quantum computers require microwave control lines running from room temperature to the mixing chamber of a dilution refrigerator. Adding more lines without preliminary thermal modeling to make predictions risks overwhelming the cooling power at each thermal stage. In this paper, we investigate the thermal load of SC-086/50-SCN-CN semi-rigid coaxial cable, which is commonly used for the control and readout lines of a superconducting quantum computer, as we increase the number of lines to a quantum processor. We investigate the makeup of the coaxial cables, verify the materials and dimensions, and experimentally measure the total thermal conductivity of a single cable as a function of the temperature from cryogenic to room temperature values. We also measure the cryogenic DC electrical resistance of the inner conductor as a function of temperature,
allowing for the calculation of active thermal loads due to Ohmic heating. Fitting this data produces a numerical thermal conductivity function used to calculate the static heat loads due to thermal transfer within the wires resulting from a temperature gradient. The resistivity data is
used to calculate active heat loads, and we use these fits in a cryogenic model of a superconducting quantum processor in a typical Bluefors XLD1000-SL dilution refrigerator, investigating how the thermal load increases with processor sizes ranging from 100 to 225 qubits. We conclude that the theoretical upper limit of the described architecture is approximately 200 qubits. However, including an engineering margin in the cooling power and the available space for microwave readout circuitry at the mixing chamber, the practical limit is approximately 140 qubits.

\end{abstract}

\pacs{Valid PACS appear here}% PACS, the Physics and Astronomy
\maketitle
\section{Introduction}
Superconducting quantum computers use qubits with energy gaps between the ground and the first excited states in the microwave regime. Frequencies in this range allow for qubit manipulation and readout using microwave components in the low-photon number regime, but also require the devices to be operated at temperatures of order 20 mK so that the thermal population of the first excited state is negligible. Commercially available dilution refrigerators achieve these temperatures, but the system must be engineered such that the total heat load coming from the outside or generated within the cryostat is lower than the cooling power available from the refrigeration system. The qubit gate operation and readout are controlled by microwave pulses generated by room-temperature electronics outside the cryostat. These pulses travel down through signaling cables that are attached at the different temperature stages of the dilution refrigerator. The cables create a static heat load due to thermal conductivity and temperature differences between stages. Any cable that carries a current also has an active load due to Ohmic heating, due to the cable resistivity, and heat loss in attenuators used to thermalize the cable at each stage \cite{clarke2008}.

Scaling the number of qubits requires an increase in the number of signal lines; however, there is a limit to the number of cables that can be accommodated for any one particular cable type. This limiting factor is usually either the thermal budget or the available space \cite{krinner2019}. More compact and dense microwave cable types are being actively investigated \cite{deshpande2019}, and other potential solutions to scaling the number of control signals have been reported, such as using optical fibers and a transducer to create RF pulses from optical signals \cite{lecocq2021}, or using cryogenic complementary metal oxide semiconductor (cryo-CMOS) circuits either for RF switching \cite{potocnik2021} or arbitrary waveform generation within the cryostat \cite{pellerano2022}. Still more solutions have been explored in recent years, including those that employ low-temperature superconducting digital circuits for control, such as superconductive rapid single flux quantum (RSFQ) and adiabatic quantum flux parametron (AQFP) circuits, both of which promise high energy efficiency \cite{krylov2024single}. Theoretical optimization-based techniques, such as those described in Zhuldassov et al. \cite{Zhuldassov2023}, can also be used, which identify the optimal operating temperatures for circuit components in a cryogenic computing system.

While all of these techniques show promise, an understanding of the thermal properties of the materials that comprise a signaling cable is necessary, and thermal conductivity and resistivity data allows for the strategic planning of wiring configurations such that the total power dissipation is minimized. This work extends the analysis and data presented in Krinner et al. \cite{krinner2019} in modeling the thermal loads resulting from the delivery of signals to superconducting quantum processors. We offer additional insight by discussing the viability of processor scaling to 100-225 qubits and by focusing our study on a widely used and commercially available small-format coaxial cable for which detailed datasets did not previously exist. The particular cable is the SC-086/50-SCN-CN semi-rigid coaxial cable used in the Bluefors High-Density Wiring (HDW). These cables have the same internal dimensions as the UT-085 cables considered by \cite{krinner2019}, with a cupronickel outer and silver-coated cupronickel inner conductor with a Polytetrafluoroethylene (PTFE) dielectric. Assemblies of these coaxial lines are used in the Bluefors XLD1000-SL range of dilution refrigerators to allow simplified wiring configuration changes.

In this work, Section~\ref{overview} describes a general overview of how we approach the thermal modeling, including the cooling powers available at different fridge stages and the details of the different control lines used to control a superconducting processor. Section~\ref{Structure} then details the material composition and cross-sectional dimensions of the inner signal line, dielectric, and outer ground of the SC-086/50-SCN-CN semi-rigid coaxial cable. In Section~\ref{thermal conductivity}, we report the measured thermal conductivities of the inner signal and outer ground sections as a function of temperature, along with fits to the data. In Section~\ref{static model}, we use the measured thermal conductivity fits with data for the thermal conductivity of the dielectric layer published by NIST \cite{marquardt2002, bradleyp.e.2006}, to develop a thermal model of the static heat load of the coaxial cable across all thermal stages spanned by the cable. We then consider the active loads due to Ohmic heating, beginning with Section~\ref{DC resistance}, in which we report the measured resistivity of the signal conductor of an SC-086/50-SCN-CN cable as a function of temperature along with fits to the data. Sections \ref{active model} and \ref{active model calc} describe the active load thermal model and calculations, respectively that use the resistivity data. The active load is determined by the system configuration and architecture, where the current in the signal line and the attenuators dissipate heat. Finally, in Section \ref{Bluefors}, we combine the static and active heat load models to simulate a complete superconducting quantum computer ranging in size from 100 to 225 qubits. 

\section{Overview of Thermal Modeling}\label{overview}
Current designs of dilution refrigerators are able to cool the sample space to under 20 mK without cryoliquids using a He3/He4 dilution unit that is precooled by a two-stage pulse tube refrigerator \cite{uhlig2002, uhlig2008}. The cables within a dilution refrigerator are thermally anchored at several stages within the fridge, whose temperatures are set by the active cooling systems. The nominal stage temperatures and estimated cooling powers for the Bluefors XLD1000-SL system, along with the cable lengths, are shown in Table~\ref{tab:cooling power}. Bluefors specifies the cooling power and base temperature only at the Mixing Chamber (MXC) stage \cite{Bluefors2025}, and the numbers for higher temperature stages are estimates based on ``heat maps" we have measured, where we position heaters on the various stages of a bare fridge and record the variation of stage temperatures in response to varied heater power. These cooling powers are indicated by an asterisk in Table~\ref{tab:cooling power}, and they are those available to cool experimental loads added to the bare refrigerator. The cooling powers in Table~\ref{tab:cooling power} are close to those reported by Krinner \cite{krinner2019} for the smaller Bluefors LD400 system with suitable scaling for the more powerful dilution unit. The main differences are at the 4K stage, where Krinner reports 1.5 W of cooling power at 4.2 K, whereas we use a figure of 0.7 W at 3.5 K. There are notable differences in the pulse tubes and fridge hardware here, but the lower cooling power is mainly attributable to the lower operating point temperature.

The dilution fridge is a coupled dynamic system, and the estimated cooling power of each fridge stage varies according to how the system is operated and optioned, depending on the specifications of the installed pulse tube coolers and the dilution units. Pulse tube technology has improved substantially over the past years, and the available cooling power is expected to increase. For this paper, we assume the system has two Cryomech PT420 pulse tubes, which were the standard configuration of the Bluefors XLD1000-SL at the time of model development. However, it should be noted that, as of this writing, the current default configuration uses two Cryomech PT425 pulse tubes with an option to replace one of the Cryomech PT425 pulse tubes with a Cryomech PT310 pulse tube \cite{Bluefors2025}. Another option is the use of three pulse tubes at the expense of reducing the number of side-loading ports for the wiring \cite{Bluefors2025}. The fridge has multiple wiring options, and here we focus on the SC-086/50-SCN-CN coax cable used in the Bluefors High-Density Wiring (HDW) option.

In this paper, we calculate and sum the static and active heat loads for each model to provide a total input heat load at each stage. We then calculate the final heat load at each fridge stage, taking into account the difference between the incoming heat load from higher temperature stages, the Ohmic heating, and the outgoing load to the next lower temperature stage. We compare this final load to the estimated cooling power of each fridge stage shown in Table~\ref{tab:cooling power}.

To model a complete superconducting processor, we take a common design that uses flux-tunable transmons as the basic qubit design, with tunable couplers between pairs of qubits that are designed for use in two-qubit gate operations \cite{yan2018}. The control and readout lines for these processors can be split into four different categories:

\begin{enumerate}
  \item Qubit XY drive lines for single-qubit operations.
  \item Qubit flux bias lines to set the qubit frequency.
  \item Tunable Coupler flux bias lines to set the coupler operating point.
  \item Readout lines to excite the readout resonator and perform a measurement on the qubit.
\end{enumerate}

The different functional groups of the microwave lines each have a different attenuator configuration to ensure the thermal noise on the signal line at the processor is at the base temperature of the fridge. All of the lines provide a static heat load because they run from room temperature to the MXC, while the active Ohmic load depends on the current in the line and the attenuator configuration.

In this model design, the qubit XY lines carry infrequent microwave pulses for single-qubit manipulation, while the flux bias line has a constant current to set the qubit frequency. The tunable couplers have a similar flux bias line with a constant current to set the operating point so that the two qubits do not interact, which is changed to enable two-qubit gate interactions. Each qubit is coupled to a resonator for readout, and the readout circuit has a read-in line coupled to several resonators and continues on to a readout line with isolators and amplifiers. A Traveling Wave Parametric Amplifier (TWPA) and a low noise semiconductor amplifier are frequently used to boost the output signal in the readout. The TWPA requires a separate microwave line to provide a microwave pump signal, and a semiconductor amplifier has twisted-pair DC wiring and produces heat that must be accounted for. 

%=============Table 1, Fridge Parameters ======================

\begin{table}[h!]
    \renewcommand*{\arraystretch}{1.4}
    \centering
    \begin{tabular}{c|c|c|c}
    \hline
    \multicolumn{4}{c}{Bluefors XLD1000-SL Dilution Refrigerator}\\
    \multicolumn{4}{c}{Estimated values configured with two PT-420 pulse tubes}\\
    \hline
    Refrigerator & Nominal & Cooling & Cable Length\\
    Stage Name & Temperature (K)& Power (W) & (m)\\
    \hline
    300K & 297 &  & \\
    \hline
    50K & 40$^{*}$ & 30$^{*}$ & 0.3053\\
    \hline
    4K & 3.5$^{*}$ & 0.7$^{*}$ & 0.3155 \\
    \hline
    Still & 1.4$^{*}$ & 7e-3$^{*}$ & 0.2775 \\
    \hline
    CP & 0.2$^{*}$ & 1e-3$^{*}$ & 0.1965 \\
    \hline
    MXC & \multirow{2}{*}{0.02} & \multirow{2}{*}{30e-6} & 0.1965 \\
  
    Below MXC &  &  & 0.1965 \\
    \hline
    \end{tabular}
    \caption{Estimated temperatures and cooling power at each refrigerator temperature stage of a  Bluefors XLD1000-SL system for a bare fridge. Cable lengths refer to the cable run from the higher temperature stage. ``Below MXC" refers to the cables run from the MXC to the package holding the quantum processor. The Below MXC cables are not considered in the static heat load as they are isothermal at the mixing chamber temperature. They do, however, contribute an active load. Bluefors specifies the cooling power and base temperature only at the MXC stage \cite{Bluefors2025}; the numbers for higher temperature stages are estimates based on our own tests of stage temperature variation in response to varied heating power, and they are indicated by an asterisk.}
    \label{tab:cooling power}
\end{table}

% ===================================

\section{Static Heat Loads}\label{static heat load}

Coaxial cables carry heat from the hotter fridge stages, and each material component has a different thermal conductivity. We model the inner and outer conductors and the dielectric layer of the coaxial cable as separate parallel thermal paths, which we then sum to produce a total static heat load. The model requires the dimensions of the material components, the fixed temperatures at each end of the cable, and the thermal conductivity of each material as a function of temperature.

While comprehensive thermal conductivity functions exist for various grades of copper and aluminum, models for certain cupronickel alloys are not generally available from cryogenic to room temperatures. In this study, we measure the thermal conductivity of the outer and inner conductors of SC-086/50-SCN-CN coaxial cable samples (Coax Co., Ltd.) as a function of temperature from cryogenic to room temperature. For the thermal conductivity of the PTFE dielectric layer, we use the published Teflon $k(T)$ function from the NIST Cryogenics Index of Material Properties database \cite{marquardt2002, bradleyp.e.2006}. While PTFE and Teflon are both ways to refer to Polytetrafluoroethylene, in the present paper, we refer to this material as ``PTFE".

The static model only considers heat transport by thermal conduction. While infrared radiation is present, the cryogenic temperatures and the low emissivity of the outer conductor of SC-086/50-SCN-CN coaxial cable ensure it is insignificant compared to the thermal conductivity \cite{krinner2019}. The only time we account for the correction due to infrared radiation is in the experimental measurement of thermal conductivity, where the Quantum Design software incorporates an estimate of the radiative losses into the measurement, and thus, we must ensure that we set the emissivity to a realistic value.

\subsection{Composition and Construction of SC-086/50-SCN-CN Coaxial Cables} \label{Structure}

%==================== Figure 1, Cross-section of coaxial cable ==========

\begin{figure*}
    \centering
    \includegraphics[width=1\textwidth]{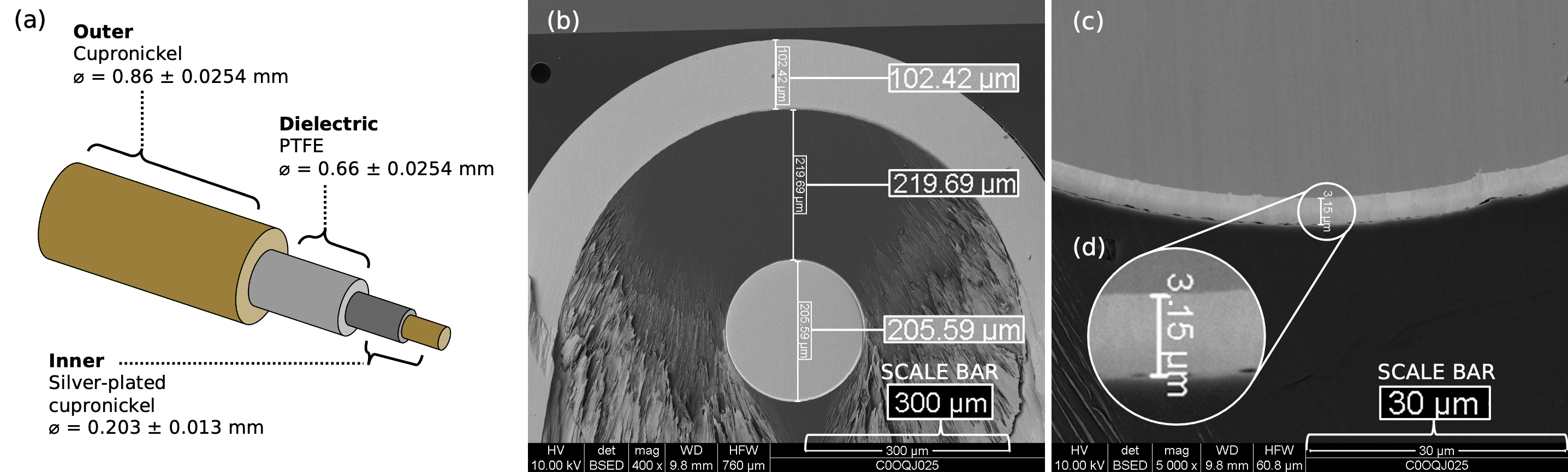}
    \caption{(a) Depiction of the layers (not to scale) of an SC-086/50-SCN-CN coaxial cable (Coax Co., Ltd.) with materials and dimensions taken from the manufacturer's website \cite{sc-08650}. (b) The SEM cross-section of an SC-086/50-SCN-CN coaxial cable sample from Coax Co., Ltd. has measured dimensions that agree with the reported dimensions in \ref{fig:cross-section}(a). Measured dimensions and scale are enlarged for readability. (c) A higher magnification image of the inner conductor of an SC-086/50-SCN-CN coax cable sample. The measured silver coating is approximately 3 $\mu m$ thick, which agrees with the nominal thickness \cite{coaxcocomm}. Measured dimensions and scale are enlarged for readability. (d) An enlarged portion of \ref{fig:cross-section}(c) to make the dimension of the silver coating readable. A larger decimal point was digitally added to the image for clarity.
    }
    \label{fig:cross-section}
\end{figure*}

% ===================================

The SC-086/50-SCN-CN cable (Coax Co., Ltd.) has a C7150 cupronickel outer conductor, a PTFE dielectric, and a silver-coated (nominal thickness= 3 - 4 $\mu$$m$) \cite{coaxcocomm} C7150 inner conductor. We detail the material specifications and dimensions in Table \ref{tab:Dimensions}, which are taken from \cite{sc-08650}. 

A diagram of the construction of the cable along with nominal dimensions is shown in Fig. \ref{fig:cross-section}(a). Cross-sectional scanning electron microscope pictures of a cable sample are shown in Figs. \ref{fig:cross-section}(b) and \ref{fig:cross-section}(c), which confirm the cable dimensions, including the thickness of the silver plating on the signal line. The composition of the cupronickel inner and outer conductors, as well as the dielectric, were measured using Energy Dispersive X-ray (EDX) analysis to determine the atomic ratios of materials. The measured atomic ratios of copper to nickel for the cupronickel alloy in both conductors, which is approximately 62.5:37.5, and the measured atomic ratio of carbon to fluorine for the dielectric, which is close to the 1:2 ratio expected from the (C$_{2}$F$_{4}$)$_{n}$ molecular structure of PTFE, are listed in Table \ref{tab:Composition}.

%==================== Table 2, Dimension of coaxial line  ==========

\begin{table}[h!]
    \renewcommand*{\arraystretch}{1.4}
    \centering
    \begin{tabular}{c|c|c|c}
    \hline
    \multicolumn{4}{c}{SC-086/50-SCN-CN Specifications}\\
    \hline
    & Diameter & Cross-Sectional &  \\
    & (mm) & Area (mm\textsuperscript{2})  & Material\\
    \hline
    Outer & 0.86 & 0.2389  & C7150\\
    \hline
    Dielectric & 0.66 & 0.3098 &  PTFE\\
    \hline
    Inner & 0.203 & 0.0324 &  Ag-Plated C7150\\
    \hline
    \end{tabular}
    \caption{Dimensional and Material Specifications of SC-086/50-SCN-CN Coaxial Cable \cite{sc-08650}. The cross-sectional areas of the material in each layer are calculated from the diameters of the different components, subtracting the area of the inner layers for the outer and dielectric layers.}
    \label{tab:Dimensions}
\end{table}

%==================== Table 3, EDX Analysis of coaxial line  ==========

\begin{table}[h!]
    \renewcommand*{\arraystretch}{1.4}
    \centering
    \begin{tabular}{c|cc|ccc}
    \hline
    \multicolumn{6}{c}{Energy Dispersive X-ray Materials Analysis}\\
    \multicolumn{6}{c}{of SC-086/50-SCN-CN Coaxial Cable}\\
    \hline
    \multicolumn{1}{c|}{} &
    \multicolumn{2}{c|}{Inner Conductor} &
    \multicolumn{1}{c}{} &
    \multicolumn{2}{c}{Outer Conductor} \\
    \hline
    Element & Wt\% & at\%& & Wt\% & at\%\\
    \hline
    C & 3.58 & 16.04 & & 3.46 & 15.68\\
    \hline
    Ni & 34.50 & 31.59 & & 34.39 & 31.59\\
    \hline
    Cu & 61.92 & 52.37 & & 62.15 & 52.73\\
    \hline
    Cu : Ni ratio & \multicolumn{2}{c|}{62.4 : 37.6} &
    \multicolumn{1}{c}{} &
    \multicolumn{2}{c}{62.5 : 37.5}\\
    \hline
    \noalign{\smallskip}
    \cline{1-3}
    \multicolumn{1}{c|}{} &
    \multicolumn{2}{c|}{Dielectric} &
    \multicolumn{1}{c}{} &
    \multicolumn{2}{c}{} \\
    \cline{1-3}
    Element & Wt\% & at\% & &\\
    \cline{1-3}
    C & 23.81 & 33.1 & &\\
    \cline{1-3}
    F & 76.19 & 66.9 & &\\
    \cline{1-3}
    C : F ratio & \multicolumn{2}{c|}{33.1 : 66.9} & &\\
    \cline{1-3}
    \end{tabular}
    \caption{Compositional analysis of the inner and outer conductors and the dielectric layer, measured by EDX on the cross-sectional sample shown in Fig. \ref{fig:cross-section}. For the inner conductor measurement, the electron probe was far away from the silver coating, which excluded silver from the analysis. The Cu:Ni ratio is approximately 62.5:37.5 in the two conductors. In the dielectric, the carbon-fluorine ratio is close to the theoretical value of 1:2 from the molecular structure (C$_{2}$F$_{4}$)$_{n}$.}
    \label{tab:Composition}
\end{table}

%==================== Figure 2, PPMS thermal conductivity ==========

\begin{figure}
    \centering
    \includegraphics[width=\columnwidth]{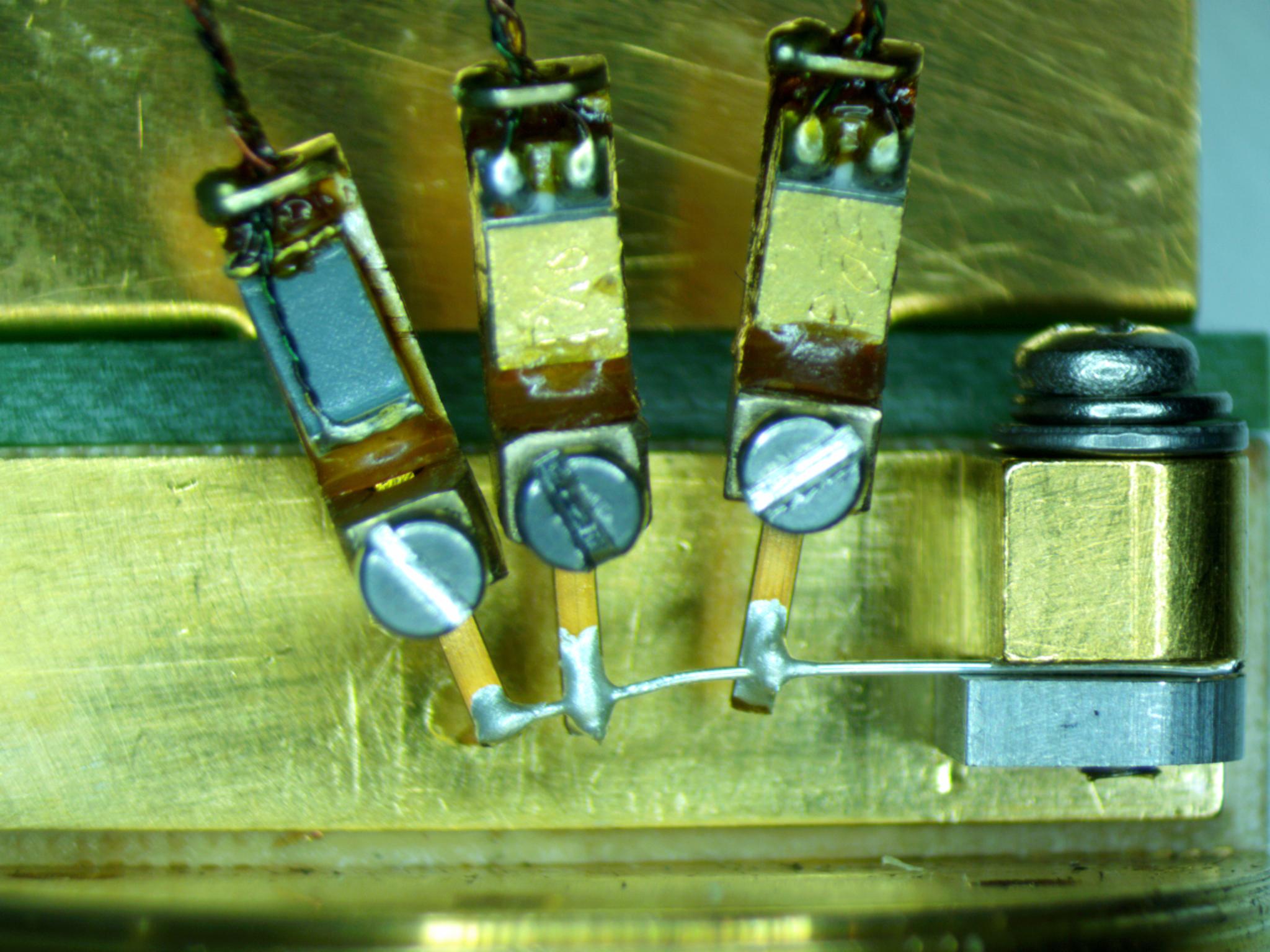}
    \caption{The cupronickel outer conductor of an SC-086/50-SCN-CN coaxial cable mounted on the PPMS TTO measurement system. From left to right, we see the sample heater, which produces a time-dependent thermal gradient across the sample, two thermometers, and then thermal anchoring of the sample to the right-hand end.
    }
    \label{fig:device diagram}
\end{figure}

% ===================================

\pagebreak
\subsection{Experimental Measurement of Thermal Conductivity}\label{thermal conductivity}

A sample of SC-086/50-SCN-CN coaxial cable was deconstructed into its three constituent components: the C7150 cupronickel outer conductor, the PTFE dielectric, and the silver-coated C7150 cupronickel inner conductor. The thermal conductivity of the outer and inner conductors were each measured using the Thermal Transport Option (TTO) in a DynaCool Physical Property Measurement System (PPMS) by Quantum Design, Inc. \cite{maldonado1992, zotero-361}. Using this method, the thermal conductance of the sample is measured and then converted to thermal conductivity based on the temperature sensor lead separation, cross-sectional area, and emissivity. 

The TTO measurement setup, which is shown in Fig. \ref{fig:device diagram}, involves the connection of an electrical heater to the free end of a sample and the thermal anchoring of the other end to a variable temperature stage. The heater is operated in a pulsed mode, and two thermometers placed along the sample measure the local temperature. Fitting the rise and fall of the temperature change versus time allows the thermal conductance to be extracted \cite{maldonado1992}. The sample emissivity is a parameter of the thermal model, which includes an estimate of the radiative heat loss.

Samples were measured over a 2 - 300 K temperature range. Inner conductor lead separation was measured to be 2.69 mm, while outer conductor lead separation was 6.96 mm. The inner conductor cross-section was estimated to be 0.0324 mm\textsuperscript{2}, given the reported 0.203 mm outer diameter of the circular inner conductor. The outer conductor cross-section was estimated to be 0.2389 mm\textsuperscript{2}, given the annular shape of the outer conductor cross-section and given the reported 0.86 mm outer diameter of the outer conductor and the reported 0.66 mm outer diameter of the dielectric layer. The cross-sectional dimensions of the cable are summarized in Table \ref{tab:Dimensions}.

The correction for radiative losses is largest at higher temperatures, even though, in this case, it remains a small correction \cite{krinner2019}. For the emissivity input for these corrections, we selected fixed emissivity values measured at 300 K from the literature, the highest temperature at which we measured thermal conductance. Although emissivity is known to vary with temperature, there is only a slight variation over the range 300 - 50 K, where the correction for radiation losses is typically significant compared to the sample thermal conductance. The inner conductor has a 3 $\mu m$ silver coating, and the emissivity is taken to be $\epsilon = $ 0.00775, from the measured emissivity at 300 K of a silver coating ($>$ 5 $\mu m$) on a polished SS304L stainless steel surface reported by Woods et al. 2014 \cite{woods2014}. The outer conductor is a cupronickel alloy of 62.5:37.5 Cu to Ni ratio. The emissivity value is taken to be $\epsilon = $ 0.0813 based on a curve fit of emissivity versus composition of cupronickel alloy reported in Gelin et al. 2004, where they refer to this emissivity as ``thermal emittance" \cite{gelin2004}.

\begin{figure}
    \centering
    \includegraphics[width=\columnwidth]{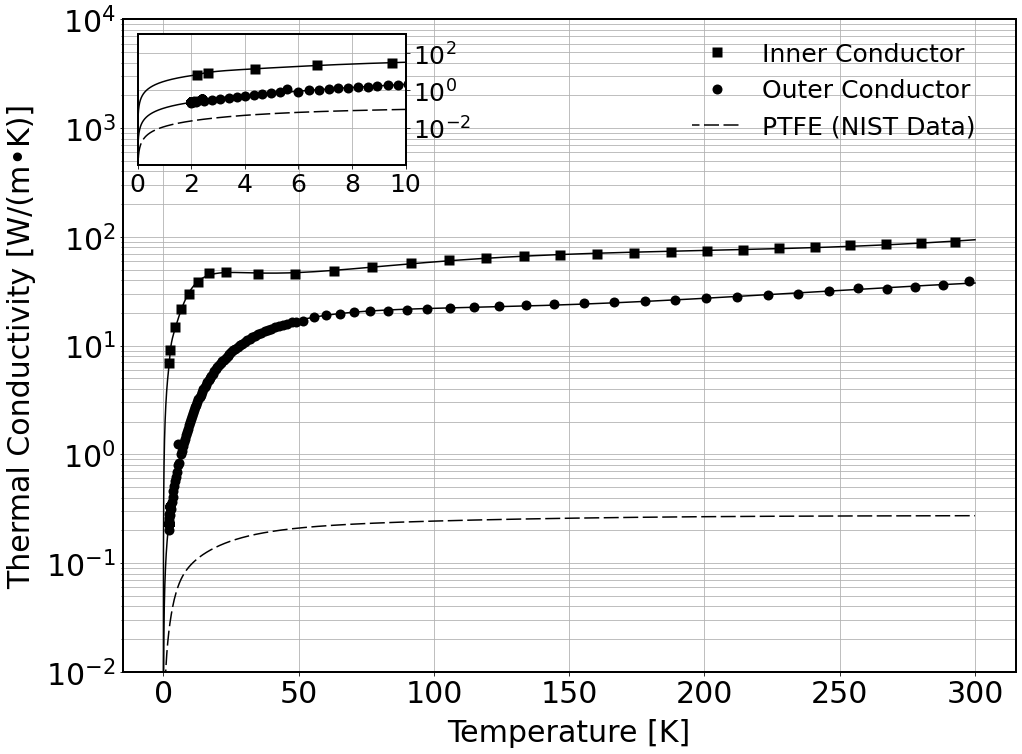}
    \caption{Thermal conductivity versus temperature for silver-coated cupronickel inner conductor, cupronickel outer conductor, and PTFE. Due to the challenge in accurately modeling the radiative correction above 300 K, the raw data was processed to exclude points above 300 K; the curves shown fit the processed data. The PTFE function displayed is the Teflon model from the NIST Cryogenics Index of Material Properties database \cite{marquardt2002, bradleyp.e.2006}.
    }
    \label{fig:k graph}
\end{figure}

Experimental thermal conductivity data for the inner and outer conductors are shown in Fig. \ref{fig:k graph}, along with a PTFE thermal conductivity curve supplied by the NIST Cryogenics Index of Material Properties database \cite{marquardt2002, bradleyp.e.2006}.

%==================== Figure 3, k(T) figure ==========
% ===================================

\subsection{Fitting the Thermal Conductivity of SC-086/50-SCN-CN Components} \label{fit}
 
The inner and outer conductor PPMS TTO thermal conductivity data are shown in Fig.~\ref{fig:k graph}. Each processed dataset was fitted using a least-squares polynomial fitting procedure to obtain a smooth $k(T)$ function valid for temperatures shown in the equation range row in Table~\ref{tab:fit coefficients}. To process the data, measurements performed above 300 K were discarded due to the challenge in accurately modeling the radiative correction above 300 K.

The $k(T)$ fit function is based on an 8th-degree Taylor series of the base-10 logarithm of temperatures $T$, a format used in the NIST cryogenic database. The utility of this particular format lies in the fact that it can handle functions that change many orders of magnitude:

\begin{equation} \label{eqn:fit eqn}
\begin{split}
k(T) = 10\hat{\mkern6mu} [a &  + b\left(\log_{10}{T}\right) + c\left(\log_{10}{T}\right)^{2}+ d\left(\log_{10}{T}\right)^{3} \\
 & + e\left(\log_{10}{T}\right)^{4} + f\left(\log_{10}{T}\right)^{5} + g\left(\log_{10}{T}\right)^{6} \\
 & + h\left(\log_{10}{T}\right)^{7} + i\left(\log_{10}{T}\right)^{8}]
\end{split}
\end{equation}
The least-squares fitting procedure determined the value of alphabetical coefficients $a$ through $i$ in equation (\ref{eqn:fit eqn}). These coefficients were used to plot the smooth inner and outer conductor thermal conductivity curves shown in Fig.~\ref{fig:k graph}. Similarly, the coefficients provided by the NIST Cryogenics database for PTFE were used to plot the thermal conductivity over the range 4 - 300 K.

%==================== Table 4, thermal conductivity fit table ==========

\begin{table}[h!]
    \renewcommand*{\arraystretch}{1.4}
    \small
    \centering
    \begin{tabular}{c|c|c|c}
    \hline
    \multicolumn{4}{c}{Thermal Conductivity Fit (W/(m$\cdot$K))} \\
    \hline
    Coefficient  & Outer & Dielectric & Inner\\
    \hline
    a  & -3.198399 & 2.7380 & -2.750003\\
    \hline
    b  & 20.49947 & -30.677 & 25.84512\\
    \hline
    c  & -66.11415 & 89.430 & -74.18405\\
    \hline
    d  & 117.6898 & -136.99 & 113.5856\\
    \hline
    e  & -121.4773 & 124.69 & -96.84387\\
    \hline
    f  & 76.21467 & -69.556 & 46.38328\\
    \hline
    g  & -28.74949 & 23.320 & -11.82451\\
    \hline
    h  & 5.984756 & -4.3135 & 1.321682\\
    \hline
    i  & -0.5266892  & 0.33829 & -0.02456645\\
    \hline
    Valid equation range& 2.0 - 297.6 K& 4 - 300 K& 2.3 - 292.6 K\\
    \hline
    \end{tabular}
    \caption{Fitting coefficients for thermal conductivity as a function of temperature. ``Inner" coefficients fit our $k(T)$ data for the silver-coated C7150 inner conductor of an SC-086/50-SCN-CN cable, and ``outer" coefficients fit our $k(T)$ data for the C7150 outer conductor for the same cable. ``Inner" and ``outer" coefficients were rounded to seven significant figures, and the full seven significant figures are required to maintain the accuracy of the fits at the high temperature end. Data points collected for the inner and outer conductors correspond to temperatures that range from approximately 2.0-297.6 K for the outer conductor and approximately 2.3-292.6 K for the inner conductor. Polynomial fits were used to calculate thermal conductivity for temperatures in the reported ranges for each material and up to 300 K. A linear extension to the origin from the data point corresponding to the lowest temperature in the set of used points was used for modeling thermal conductivities at low temperatures. PTFE coefficients are sourced from the NIST Cryogenics Index of Material Properties database ~\cite{marquardt2002, bradleyp.e.2006}.}
    \label{tab:fit coefficients}
\end{table}

% =================================== 

Since our work requires thermal conductivities down to 20 mK, we need to extend the fits below the lower end of the experimental data and the validity of the power series function. Here we assume that the thermal conductivity of materials goes to zero at zero temperature, and we extend the thermal conductivity curve using a straight line that connects the lowest-temperature experimental $k$ data point down to the origin. This procedure likely overestimates the thermal conductivity in this temperature range, as the slope of the experimental data near the lowest temperatures is larger than the slope of the straight-line approximation.

The fitting coefficients $a$ through $i$, as determined by the least-squares polynomial fit for the outer and inner conductors, are listed in Table ~\ref{tab:fit coefficients}. The full seven significant figures are required to maintain the accuracy of the fits at the high temperature end. We used these coefficients to graph the solid black curves of best fit for the inner and outer conductor data in Fig. \ref{fig:k graph}. We plotted the NIST PTFE curve using the coefficients available from the NIST Cryogenics Index of Material Properties database \cite{marquardt2002, bradleyp.e.2006}.

\subsection{Method of Calculating Static Heat Load}\label{static model}

We model the coaxial cable as three parallel thermal conductances, one for each cable component: the outer conductor, dielectric, and inner conductor. The total thermal conductance of the cable is then the sum of the conductance of each component.

While our model does not assume that the temperature profile along any one cable component exactly matches another, it ignores any radial heat transfer across the cable. We justify this simplification by pointing out that the construction of the cable is such that the two conductors are separated by a low thermal conductivity dielectric, an arrangement that limits the radial heat flow. Radial heat flow becomes more relevant in the case of significant Ohmic heating due to a large current flow in the cable, where the inner and outer conductors might have very different temperatures. 

If the cable components have a constant cross-sectional area and length, the static heat load associated with a given cable component can be calculated by numerically integrating over the thermal conductance for that material as a function of temperature using a fit to the function such as those discussed in Section \ref{fit}:

\begin{equation}
\label{eqn:intro_eqn}
 q_{component} = \frac{A}{L}\int_{T_L}^{T_H} k(T) \,dT 
\end{equation}
where $q_{component}$ is the static heat load of the cable component, $T_L$ and $T_H$ are the low and the high temperatures spanning either end of a section of cable, $A$ is the cross-sectional area, $L$ is the length of the cable section, and $k(T)$ is the thermal conductivity of the component material as a function of temperature $T$.

We sum the static heat load associated with each cable component - the outer conductor, dielectric, and inner layers - to determine the total static heat load for a single cable:

\begin{equation}
\label{eqn:intro_eqn_2}
q_{cable} = \sum_{}^{}  q_{component} = q_{outer} + q_{dielectric} + q_{inner}
\end{equation}
where $q_{outer}$, $q_{dielectric}$, and $q_{inner}$ respectively refer to each static heat load associated with the components of a coaxial cable. When thermally modeling a line in a dilution fridge, we calculate the static heat load for each temperature stage as the incoming heat from the higher temperature stage minus the heat load being removed to a lower temperature stage.

The calculated static heat loads of a single SC-086/50-SCN-CN cable in the Bluefors  XLD1000-SL system, with the stage temperatures and line lengths given in Table~\ref{tab:cooling power}, are shown in Table~\ref{tab:single_cable_static_heatload}.

%============= Table 5, Static Heat Load of a single cable ======================

\begin{table}[h!]
    \renewcommand*{\arraystretch}{1.4}
    \centering
    \begin{tabular}{c|c|c}
    \hline
    \multicolumn{3}{c}{Calculated Static Heat load of }\\
    \multicolumn{3}{c}{SC-086/50-SCN-CN coaxial cables}\\
    \hline
    & Single Cable & $N$ = 1008 Cables\\
    Fridge Stage & Static Heat Load (W) & Static Heat Load (W)\\
    \hline
    50K & 6.983e-03 & 7.039\\
    \hline
    4K & 3.550e-04 & 3.579e-01\\
    \hline
    Still & 1.963e-06 & 1.978e-03\\
    \hline
    CP & 6.877e-07 & 6.932e-04\\
    \hline
    MXC & 1.448e-08 & 1.460e-05\\
    \hline
    \end{tabular}
    \caption{The calculated static heat load of a single SC-086/50-SCN-CN coax cable, and the maximum loading of $N$ = 1008 cables, in a Bluefors  XLD1000-SL system at each fridge stage.}
    \label{tab:single_cable_static_heatload}
\end{table}

We calculate $q_{stage}$, the total static heat load of $N$ cables into a dilution refrigerator temperature stage, for each refrigerator temperature stage by multiplying $q_{cable}$ by the number of cables $N$.

\begin{figure}
    \centering
    \includegraphics[width=\columnwidth]{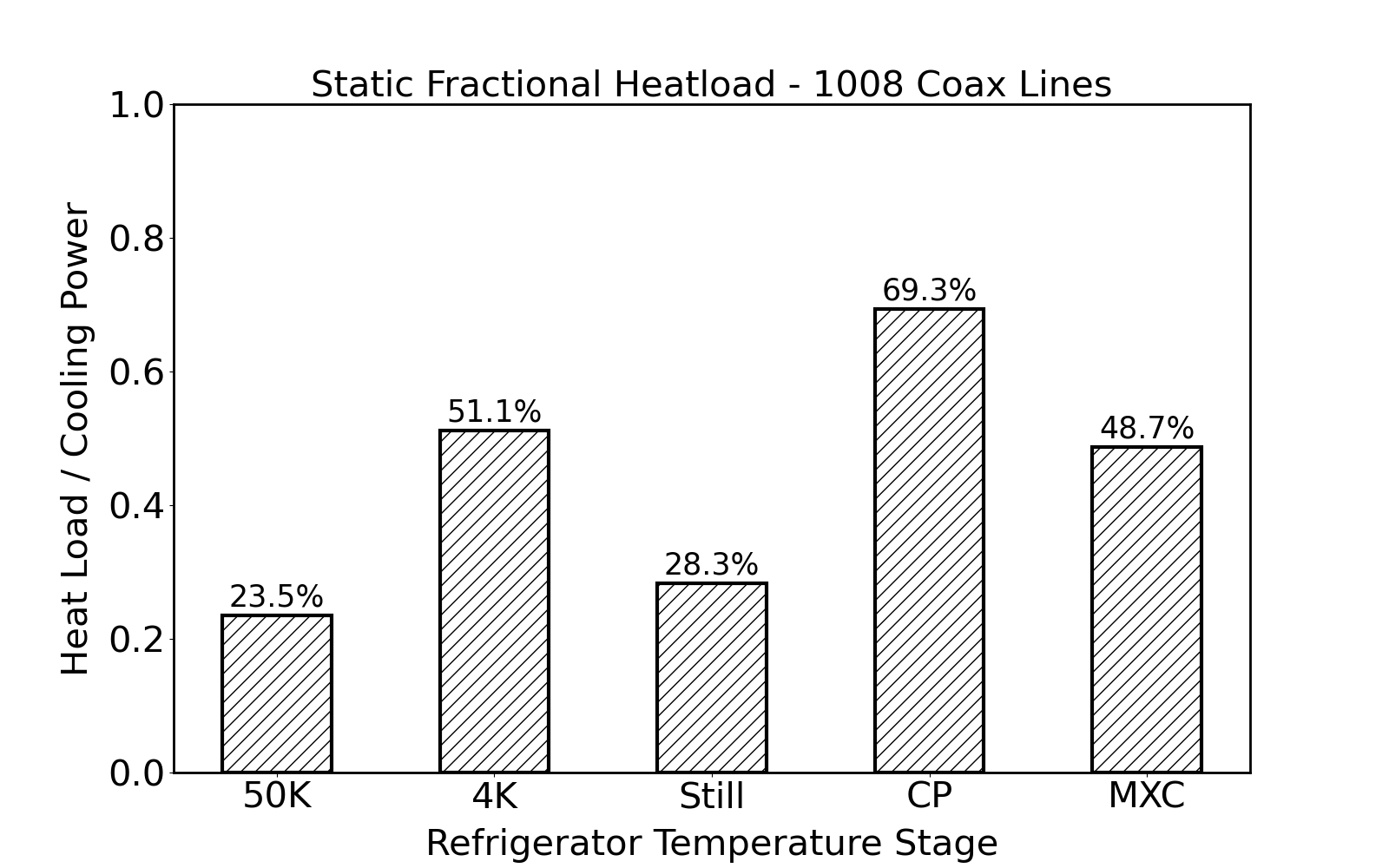 }
    \caption{The fraction of the estimated cooling powers at each stage taken up by static heat loads of a Bluefors  XLD1000-SL system outfitted with the maximum number of cables ($N = 1008$). The different refrigerator temperature stages on the x-axis are, as before, denoted as follows: ``50K" for the 50K plate, ``4K" for the 4K plate, ``Still" for the still stage, ``CP" for the cold plate, and ``MXC" for the mixing chamber.
    }
    \label{fig:SHL graph}
\end{figure}

The maximum number of Bluefors HDW SC-086/50-SCN-CN cables currently available in a Bluefors XLD1000-SL system is $N$ = 1008 \cite{bluefors2021}. We calculated the fraction of the available cooling power at each stage by taking the ratio between $q_{stage}$ and the cooling power at each stage as listed in Table~\ref{tab:cooling power}. The results are plotted as a bar chart in Fig. \ref{fig:SHL graph} and show that the largest fractional heat load is surprisingly not at the MXC but rather at the CP, where the cables take up 69.3\% of the available cooling power. We had expected that the MXC stage would have the highest fractional heat load since it has the smallest cooling power by a large margin, but upon further reflection, it becomes clear that the architecture that we consider for this paper results in a higher proportional load on the CP than on the MXC. Specifically, the CP has a shorter length of cable run from the higher stage compared to the stages above it, increasing static load from above. Further, the CP has a relatively large temperature differential of 1.4 K to 0.2 K compared with the CP to MXC.

There is a sufficient engineering margin for the MXC to reach the base temperature. Using superconducting cables below the 4K stage in place of SC-086/50-SCN-CN would reduce this load dramatically.

%==================== Figure 4, Static heatload of 1008 cables ================

% =================================== 

\section{Active Heat Loads}\label{active model}

All cables carrying an appreciable current produce an Ohmic heat load within the cryostat. In addition to the resistance of the cable, the heat load in the attenuators used to thermalize the signal line must also be accounted for. Thermalization of the signal line is required to reduce the blackbody radiation present in the cables to a level corresponding to a thermal photon occupation number, typically taken to be $\approx$ 10$^{-3}$, requiring a total attenuation of at least 60 dB \cite{krinner2019}. The effects of thermally generated electrical noise in cables and components on qubit dynamics are examined in \cite{vaaranta2022, simbierowicz2024}. Infrared filtering elements are also included in the coax line to prevent radiation from higher temperature stages from propagating through the dielectric layer. 

The active load in a coaxial cable is distributed along the cable, but the heat flows predominantly to the lower-temperature end of the cable, where it adds to the static heat load. Due to its smaller cross-sectional area, the inner conductor of the cable has a significantly higher resistance compared with the outer conductor; in our active load model, we only consider the Ohmic heating in the inner conductor. The return currents flow through the much lower-resistance outer conductor and the fridge mounting hardware, which is electrically connected to the grounds of the cables \cite{krinner2019}. In the next section, we measure the DC electrical resistivity of the inner conductor as a function of temperature, which we later use as an input to the active thermal model.

A superconducting quantum computer typically has other sources of heat within the cryostat beyond the active loads in the cables and attenuators, such as semiconductor amplifiers at the 4K stage and TWPAs at the mixing chamber with a microwave pump signal that is terminated in a 50 $\Omega$ load. The active load thermal models depend on the system configuration and the current in the signal lines. Due to the limited available cooling power, the fraction of available MXC cooling power taken up by the processor is particularly sensitive to the resistance of the cables and the connections to the package holding the quantum processor at the mixing chamber stage. In addition to the limited cooling power, generating heat near the quantum processor creates potential problems as the processor package could become locally hotter than its surroundings unless the processor has an adequate thermal connection to the mixing chamber, something that the available thermometers might not pick up. The static heat load models do not need to include the processor connections at the mixing chamber as all of those cables are isothermal to the MXC plate, and hence, there is no static load below the MXC.

\subsection{Experimental Measurement of DC Resistivity}\label{DC resistance}

We measured the DC electrical resistivity of the inner conductor of an SC-086/50-SCN-CN cable between room temperature and 3.8 K as determined by a Cernox temperature sensor. We mounted a 0.254 m (10$^{\:\prime\prime}$) length of SC-086/50-SCN-CN coaxial cable on the 4K stage of a dilution refrigerator by connecting the two ends to two SMA feedthroughs on a KF-40 flange. This guarantees the efficient thermalization of the cable, the same way the coaxial cables are anchored to the flanges in conventional coaxial builds in dilution refrigerators. We looped the cable back on itself so that both ends were attached to the 4K stage, ensuring that the length of the cable was isothermal to the 4K plate. We then measured the resistance when cooling down and warming up the fridge using a four-terminal setup through a micro-D connector compatible with the DC looms inside the dilution refrigerators. This resistance data was then scaled by the known length and cross-sectional area to give the resistivity, and we fitted the data with the eighth-order polynomial shown in equation \ref{eqn:resistivity fit eqn}.

%==================== Figure 5, rho(T) figure ==========
\begin{figure}
    \centering
    \includegraphics[width=\columnwidth]{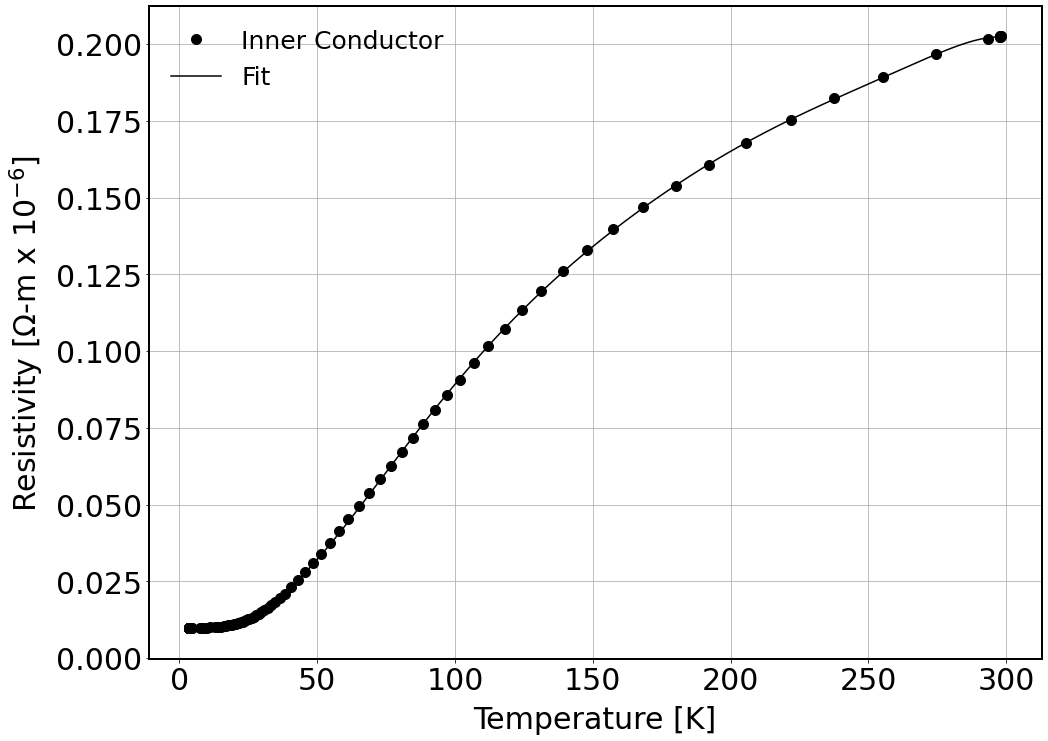}
    \caption{DC electrical resistivity versus temperature for the silver-coated cupronickel inner conductor. The data was scaled by the cross-sectional area and length of the cable to give resistivity values.
    }
    \label{fig:resistivity graph}
\end{figure}
% ===================================

%==================== Table 6, resistivity fit table ==========

\begin{table}[h!]
    \renewcommand*{\arraystretch}{1.4}
    \centering
    \begin{tabular}{c|c}
    \hline
    \multicolumn{2}{c}{DC Electrical Resistivity Fit ($\Omega\cdot$m)}\\
    \hline
    Coefficient & Inner Conductor\\
    \hline
    a &  -8.327474 \\
    \hline
    b &  10.01214 \\
    \hline
    c &  -52.83315 \\
    \hline
    d &  122.5470 \\
    \hline
    e &  -152.7599 \\
    \hline
    f &  109.0327 \\
    \hline
    g &  -44.41614 \\
    \hline
    h &  9.598158 \\
    \hline
    i &  -0.8539285 \\
    \hline
    Valid equation range & 3.8 - 300 K\\
    \hline
    \end{tabular}
    \caption{DC electrical resistivity fit parameters for the inner conductor. The coefficients were rounded to seven significant figures, and the full seven significant figures are required to maintain the accuracy of the fits at the high temperature end. We take the resistivity at temperatures below 3.8 K to be a constant of $\rho = \text{9.928e-9} \: \Omega \text{-m}$.}
    \label{tab:resistivity fit parameters}
\end{table}

The data and fit are shown in Fig.~\ref{fig:resistivity graph}, and the fit coefficients are shown in table \ref{tab:resistivity fit parameters}. The eight-order fit is in the same functional form as that used for thermal conductivity, i.e.,

\begin{equation} \label{eqn:resistivity fit eqn}
\begin{split}
\rho(T) = 10\hat{\mkern6mu} [a &  + b\left(\log_{10}{T}\right) + c\left(\log_{10}{T}\right)^{2}+ d\left(\log_{10}{T}\right)^{3} \\
 & + e\left(\log_{10}{T}\right)^{4} + f\left(\log_{10}{T}\right)^{5} + g\left(\log_{10}{T}\right)^{6} \\
 & + h\left(\log_{10}{T}\right)^{7} + i\left(\log_{10}{T}\right)^{8}]
\end{split}
\end{equation}

The resistivity approached a constant value below approximately 10 K, the temperature below which the resistivity is dominated by impurity scattering in the metal, and we observe that this value remains constant down to the base temperature of the fridge when the cable is moved to the MXC. Below the fitting temperature range of T = 3.8 K, we use a constant resistivity value of $\rho =$ 9.928e-9 $ \Omega\text{-}m$. We note that RF resistivity is dominated by the skin effect in the silver coating; thus, actual values differ from the measured DC resistivity reported here.

\subsection{Active Load Thermal Model}\label{active model}
A substantial portion of active heating within a superconducting quantum computer comes from lines with a significant current for long periods, such as flux bias control lines. Meanwhile, lines carrying low, infrequent, short-duration pulses, e.g., one-qubit gate pulses on the XY control lines and readout pulses on the read-in and read-out lines, contribute little to the thermal loads. Flux-tunable qubits require a constant offset to set the detuning of their frequency to neighboring qubits and to compensate for both fabrication spread in the Josephson junction, which leads to an offset in Josephson inductance, and any stray magnetic flux in the SQUID loop. Tunable couplers require a flux bias current to set the operating point where the coupling between adjacent qubits is zero. For the control lines in the active model, we consider only the tunable coupler and qubit flux bias signal line currents. 

The mean value of flux bias currents for qubits and tunable couplers in a system depends on the design, and measured values should be used to model a real system. For this work, we estimate that both the qubit flux bias and tunable coupler flux bias lines carry a DC current of I = 0.4 mA. We arrived at this estimate from the mutual inductance of $M = \partial \Phi / \partial I = 0.5 \Phi_{0}$/mA between the flux bias line and the qubit, where $\Phi_{0}$ is the superconducting flux quantum, as well as from a required flux amplitude of $\pm 0.2 \Phi_{0}$ \cite{krinner2019}. 

\begin{figure}
    \centering
    \includegraphics[width=2.5in]{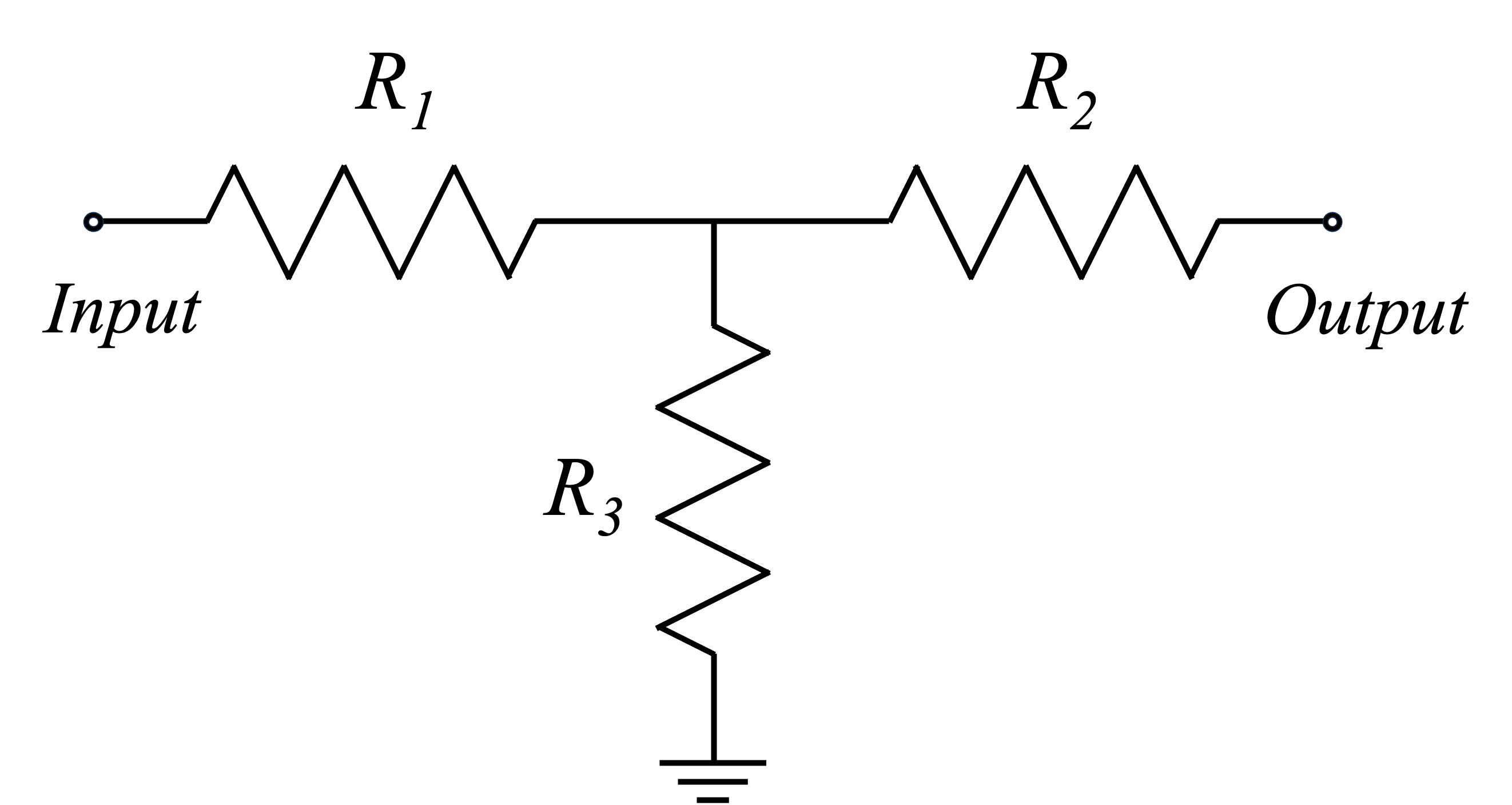}
    \caption{Circuit diagram for a T-Pad attenuator. Matching the characteristic impedance of the input and output, along with the required attenuation, allows the three resistor values to be determined, which in this case gives $R_1 = R_2 = 40.91\: \Omega$ and $R_3 = 10.1 \: \Omega$ for 20 dB of attenuation.
    }
    \label{fig:t-pad}
\end{figure}

The qubit and tunable coupler flux bias line are configured with attenuation at particular fridge stages to ensure the inner signal line is well thermalized to the cooling system \cite{yeh2017}. Each attenuator dissipates power in the resistors that must be included, feeding only a fraction of the input current to the control line below the attenuator. For this model, we assume an attenuator configuration of a single 20 dB attenuator at the 4K stage of the dilution fridge \cite{krinner2019} using a ``T-pad" attenuator shown in Fig. \ref{fig:t-pad}. Matching the 50 $\Omega$ characteristic impedance of the lines along with the required attenuation allows the three resistor values to be determined, which in this case gives $R_1 = R_2 = 40.91\: \Omega$ and $R_3 = 10.1 \: \Omega$ for 20 dB of attenuation.  The microwave calibration of attenuators in a typical signal line at cryogenic temperatures has been reported in \cite{simbierowicz2022}.

%==================== Figure 6, T-pad attenuator ==========

% ===================================

Since the output current is smaller than the input current at an attenuator, it is necessary to start with the current required at the lowest temperature stage and then calculate the input current required at each successive stage to produce the desired output current at each attenuator. We can then use these current calculations, along with measures of cable resistivity and the known resistors in the attenuator, to calculate the active heat load in each line. The assumptions on line currents and attenuator configuration are summarized in Table \ref{tab:active load currents}.

% =============== Table 7, Active Heat load control line assumptions ==================== 
\begin{table}[h!]
    \renewcommand*{\arraystretch}{1.4}
    \centering
    \begin{tabular}{c|c|c}
    \hline
    \multicolumn{3}{c}{Active Heat Load Line Currents and Attenuation}\\
    \hline
    \multicolumn{3}{c}{Qubits}\\
    \hline
    Line Type & Attenuation & Current (mA) \\
    \hline
    XY Drive &  * & 0 \\
    \hline
    Flux Bias &  20 dB 4K & 0.4 at MXC \\
    \hline
    \multicolumn{3}{c}{Tunable Couplers}\\
    \hline
    Line Type & Attenuation & Current (mA) \\
    \hline
    Flux Bias & 20 dB 4K & 0.4 at MXC \\
    \hline
    \multicolumn{3}{c}{Readout Circuit - 6 way multiplexed}\\
    \hline
    Line Type & Attenuation & Current (mA)\\
    \hline
    Read In &  * & 0\\
    \hline
    Read Out & 0 dB & 0\\
    \hline
    \end{tabular}
    \caption{A summary of the signal lines wherein the attenuation configuration and current in each of the various lines are shown; these can also be found in Fig.~\ref{fig:model_system}. An asterisk in the attenuation column indicates that while that line does have attenuators, they are irrelevant for the active load model as the line carries negligible current.}
    \label{tab:active load currents}
\end{table}

%=================================================================

% =============== Table 8, Active Heat load - TWPA & LNA assumptions ==================== 

\begin{table}[h!]
    \renewcommand*{\arraystretch}{1.4}
    \centering
    \begin{tabular}{c|c}
    \hline
    \multicolumn{2}{c}{Other Active Heat Loads}\\
    \hline
    \multicolumn{2}{c}{TWPA Pump}\\
    \hline
    Power at TWPA & -60 dBm \\
    \hline
    Power at directional coupler & -40 dBm \\
    \hline
    RMS current into 50 $\Omega$ load & 0.044 mA  \\
    \hline
     & 10 dB at 4K\\
     TWPA Pump Attenuation & 10 dB at Still\\
     & 10 dB at CP\\
    \hline
    \multicolumn{2}{c}{LNA Power}\\
    \hline
    Dissipated at 4K Stage & 7.8 mW \\
    \hline
    \end{tabular}
    \caption{A summary of other active heat loads within the cryostat. Each read-out line has a TWPA with its own pump line and an LNA at the 4K stage. We estimate the RMS current in the TWPA pump line to be the current required to dissipate the TWPA pump power at the input of the 20 dB directional coupler into a 50 $\Omega$ load.}
    \label{tab:active load other}
\end{table}

% =================================== 

Other heat sources within the cryostat must be added to the total heat loads. TWPAs \cite{macklin2015} used in the readout circuit at the MXC stage require an RF pump signal that terminates at a 50 Ohm load. We assume a typical TWPA pump drive of -60 dBm \cite{Bluefors-TWPA-2024}, fed by a 20 dB directional coupler. The TWPA pump RF power at the input to the directional coupler is therefore -40 dBm, and the vast majority of that power goes straight through and is dissipated in the 50 Ohm load. The power dissipation allows the RMS current in the pump line at the input to the directional coupler to be calculated to be I = 0.044 mA. Another heat source is from the semiconductor Low Noise Amplifiers (LNAs) that are used in the readout circuit at the 4K stage. A typical LNA that operates over the 4-8 GHz band commonly used in readout dissipates 7.8 mW of power at the 4K stage \cite{zotero-555}. The assumptions on TWPA pump and LNA loads are summarized in Table \ref{tab:active load other}.

\subsection{Method of Calculating Active Heat Load}\label{active model calc}
We calculate the active heat load by numerically integrating the resistance value along a cable multiplied by the square of the current. At any stage with an attenuator, the heat load due to the attenuator is given by the sum of the heat loads in the three resistors making up the T-pad attenuator, taking into account the different currents flowing in each. We assume that the total active heat load flows to the lower fridge stage it is thermally anchored to. Thus, the load quoted for each stage is the active load from the cable above it and the load from the attenuator, if one is installed.

The output of the thermal model is a set of thermal heat loads at each fridge stage, which the cooling system needs to handle. This analysis assumes the available cooling power and the operational temperature of fridge stages are fixed, while in reality, these operating points are coupled, which a complete model of a dilution refrigerator could capture.

\subsection{Active Load Example in a Bluefors  XLD1000-SL Dilution Refrigerator}\label{active example}
For this example, we take a typical attenuation configuration with 20 dB attenuation at the 4K plate of the fridge \cite{krinner2019} and the line lengths from a Bluefors XLD1000-SL system detailed in Table \ref{tab:cooling power}. Using the experimentally determined cable resistivity and a flux bias current of I = 0.4 mA, the calculated active heat loads of a single SC-086/50-SCN-CN cable in the Bluefors  XLD1000-SL system are shown in Table~\ref{tab:single_cable_active_heatload}.

%============= Table 9, Active heat load of a single flux bias line ======================

\begin{table}[h!]
    \renewcommand*{\arraystretch}{1.4}
    \centering
    \begin{tabular}{c|c|c|c|c}
    \hline
    \multicolumn{5}{c}{Calculated Active Heat load of a Qubit Flux Line}\\
    \multicolumn{5}{c}{Single SC-086/50-SCN-CN coaxial cable}\\
    \hline
     \multicolumn{3}{c|}{Input Parameters} &  \multicolumn{2}{c}{Heat Load} \\
     \hline
    Fridge & Attenuation & Current In & Coax & Attenuator\\
     Stage & (dB) & (mA) & (W) & (W) \\
    \hline
    50K & - & 2.029 & 4.157e-6 & - \\
    \hline
    4K & 20 & 2.029 & 6.903e-7 & 2.018e-4\\
    \hline
    Still & - & 0.400 & 1.515e-8 & - \\
    \hline
    CP & - & 0.400 & 1.120e-8 & - \\
    \hline
    MXC & - & 0.400 & 1.084e-8 & -  \\
    \hline
    \end{tabular}
    \caption{The calculated active heat load of a simulated qubit flux bias line in a Bluefors  XLD1000-SL system. This single SC-086/50-SCN-CN coaxial cable carries 0.4 mA of current at the MXC with 20 dB of attenuation installed at the 4K stage.}
    \label{tab:single_cable_active_heatload}
\end{table}

% =================================== 

%==================== Figure 7, Model System ==========

\begin{figure*}
    \centering
    \includegraphics[width=1\textwidth]{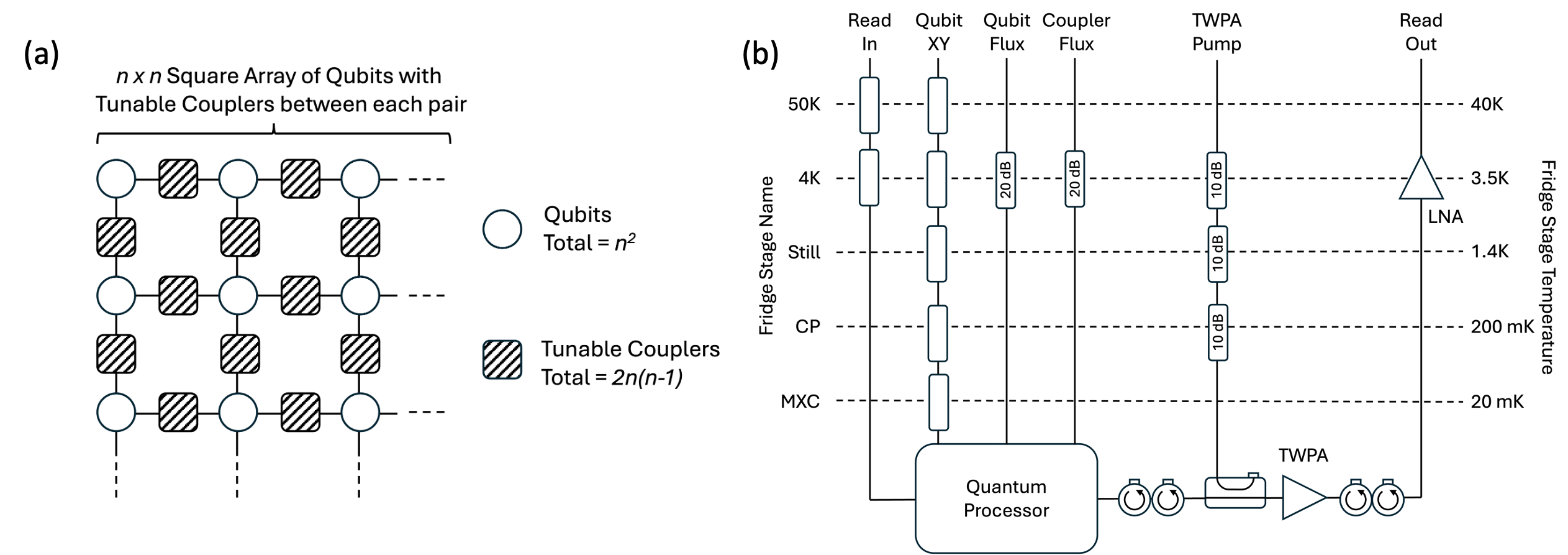}
    \caption{(a) Diagram of the layout of the model systems considered in this thermal model, which consists of a square array of $n \times n$ flux-tunable transmons and a tunable coupler between each pair of transmons. Each model has $n^{2}$ transmons and $2n \left( n-1 \right)$ tunable couplers. (b) The fridge wiring diagram of the model system, in which the assumed attenuator configuration for each type of line at each fridge stage and the readout chain electronics, including TWPAs and LNAs at the MXC and 4K stages, respectively, are shown. Note that the attenuators in the Read-In and Qubit XY lines are shown without values because they are irrelevant to the thermal model.
    }
    \label{fig:model_system}
\end{figure*}

% ===================================

%============= Table 10, Model Quantum Processors ======================

\begin{table*}[t]
    \renewcommand*{\arraystretch}{1.4}
    \centering
    \begin{tabular}{c|cc|c|cccc}
    \hline
    \multicolumn{8}{c}{Model Quantum Processor Sizes }\\
    \hline
     &  \multicolumn{2}{c|}{Components} & Circuits & \multicolumn{4}{c}{Control Lines}  \\
     \hline
    Array Size & Qubits & Tunable Couplers & Readout & Qubits & Tunable Couplers & Readout & Total \\
    $n \times n$ & $n^{2}$ & $2n \left( n-1 \right)$ & $ceil \left( n^{2}/6 \right)$ & $2n^{2}$ & $2n \left( n-1 \right)$ & $3 \: ceil \left( n^{2}/6 \right)$ & \\
    \hline
    10 $\times$ 10 & 100 & 180 & 17 & 200 & 180 & 51 & 431\\
    \hline
    11 $\times$ 11 & 121 & 220 & 21 & 242 & 220 & 63 & 525\\
    \hline
    12 $\times$ 12 & 144 & 264 & 24 & 288 & 264 & 72 & 624 \\
    \hline
    13 $\times$ 13 & 169 & 312 & 29 & 338 & 312 & 87 & 737\\
    \hline
    14 $\times$ 14 & 196 & 364 & 33 & 392 & 364 & 99 & 855 \\
    \hline
    15 $\times$ 15 & 225 & 420 & 38 & 450 & 420 & 114 & 984\\
    \hline
    \end{tabular}
    \caption{The sizes of the different model quantum processors considered. We assume each processor to be an $n \times n$ square array containing $n^{2}$ qubits, with tunable couplers between each pair of qubits. Each qubit requires two control lines: the XY drive and flux bias lines, while the tunable couplers require a single flux bias line each. The readout circuit requires three lines - a read-in line, a read-out line, and one line for the TWPA pump - and we assume that it is six-way multiplexed. Finally, each thermal model assumes that only these lines are present and there are no unused lines in the system.}
    \label{tab:model_processor_sizes}
\end{table*}

% ===================================

\section{Application of the Thermal Model to a Bluefors XLD1000-SL Dilution Refrigerator}\label{Bluefors}

In this section, we estimate the full thermal load of different-sized quantum computers with between 100 and 225 qubits installed in a Bluefors XLD1000-SL dilution refrigerator. For this estimate, we do not use the full number of cables the fridge can support. Instead, we calculate estimates for real systems using only those signaling cables required to control a particular size of processor.

The model quantum processors are shown in Fig. \ref{fig:model_system}(a) and consist of a square array of $n \times n$ flux tunable transmon qubits, where $n = 10, 11, ..., 15$, with tunable couplers between each pair \cite{yan2018}. Each model quantum processor has $n^{2}$ qubits, and $2n \left( n-1 \right)$ tunable couplers. We assume that each transmon requires two control lines - a flux bias line and an XY drive line - and that they are also each coupled to their own output resonator. Each tunable coupler has its own flux bias line. Model processors with an array of up to 15 $\times$ 15 = 225 qubits can be accommodated in the fridge, since a 16 $\times$ 16 array with 256 qubits would require a total of 1121 lines, which is beyond the capacity of $N$ = 1008 lines available.

%==================== Figure 8, Heatload results ==========

\begin{figure*}[t]
    \centering
    \includegraphics[width=1\textwidth]{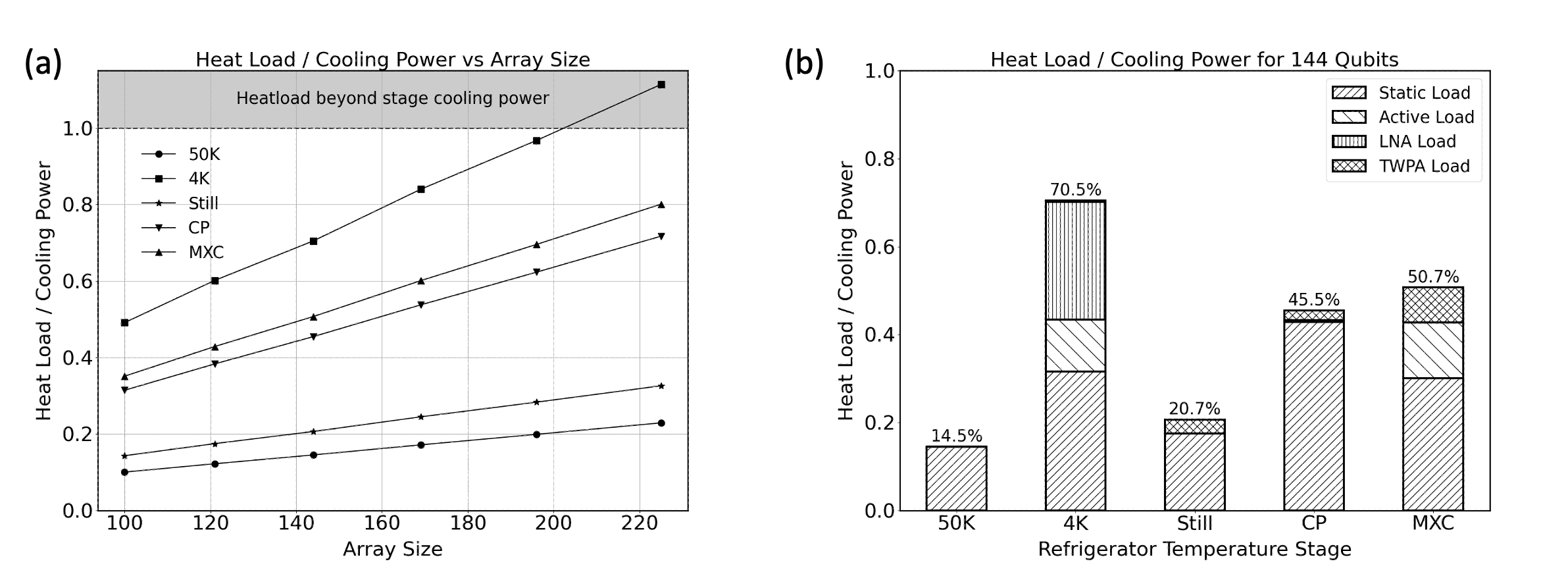}
    \caption{(a) 
    The fraction of the estimated cooling powers at the different fridge stages used by the total heat loads for model quantum processor configurations of various sizes. (b) An example of the fraction of the estimated cooling powers used at each stage for a 144-qubit processor. The LNA heatload only appears at the 4K stage, where it is situated. There are TWPA loads at the MXC due to the 50 Ohm termination resistor on the TWPA pump line, and at the 4K, Still, and CP stages due to attenuators in the TWPA pump line at those stages. The TWPA loads at the 50K and 4K stages are small, even at the 4K stage with an attenuator, and cannot be easily seen in the bar chart. Similarly, the active loads at the 50K, Still, and CP are very small fractions and cannot be easily seen in the bar chart. There are no attenuators in the flux bias lines at those stages, and the active load is due only to the cable resistance. }
    \label{fig:Total Heat load graphs}
\end{figure*}

% ===================================

Each readout circuit has a read-in line to the on-chip resonators that continues to a readout line with isolators, TWPA \cite{obrien2014, macklin2015, feng2020} fed by a pump line coming into a directional coupler, and a semiconductor LNA at the 4K stage. The readout chain assumes six-way multiplexing of the readout resonators onto $ceil \left( n^{2}/6 \right)$ separate readout circuits, where $ceil$ is the ceiling function, which rounds a fraction up to the next integer. Each of the readout circuits requires three microwave lines: one read-in, one read-out, and one line for the TWPA pump. When $n^{2}$ is not a multiple of six, one readout circuit has fewer than six qubits on it. Real designs of processors usually pick the number of qubits to be a multiple of the readout multiplexing to simplify routing, whereas these model processors are just selected to illustrate the scaling of the thermal load. The fridge wiring diagram of the model system with the assumed attenuator configuration for each type of line at each fridge stage and the specifics of the readout chain electronics, including TWPAs and LNAs at the MXC and 4K stages, respectively, are shown in Fig. \ref{fig:model_system}(b).

The total number of lines for each component in the different-sized models is shown in Table \ref{tab:model_processor_sizes}. Each thermal model assumes that only these lines are present and there are no unused lines in the system. Note that practical wiring configurations usually come in groups of cables determined by the feedthroughs at the flanges, so a real configuration may well include further static heat loads not considered here.

In Fig.~\ref{fig:Total Heat load graphs}(a), we show how the total heat loads as a fraction of the available cooling power at each stage vary with increasing processor size, using the estimated cooling powers at each stage of a Bluefors  XLD1000-SL dilution refrigerator shown in Table~\ref{tab:cooling power}. Fig.~\ref{fig:Total Heat load graphs}(b) presents a typical simulation result for 144 qubits, where the origin of the heat loads at each stage is shown in a stacked bar graph. In principle, the system has the ability to cool to base temperature provided all stages have less heat load than their nominal cooling power. 

Surprisingly, the largest fractional heat load appears at the 4K stage. In most dilution fridge experiments, the heat load at the MXC stage is the most critical because of the very limited cooling power available there. In this case, the architecture we have assumed puts a large heat load on the 4K stage due primarily to the static heat load down the cables from the 50K stage and the presence of a large number of low noise amplifiers. For a Bluefors XLD1000-SL dilution refrigerator using HDW wiring, the theoretical limit of the system using all the available cooling power is approximately 200 qubits. However, with engineering margin in the cooling power and the available space for microwave readout circuitry at the mixing chamber, the practical limit is approximately 140 qubits.

We caution readers that the model does not capture some additional heat loads that change the results. In particular, the active heat load due to the resistance of the package that holds the processor is not considered here. Other heat loads, such as the presence of a long-life cold trap, can add further static loads to the 50K and 4K stages. As such, this model only captures a general overview, and we use it to demonstrate a system with sufficient engineering margin that could be used as a starting point for a real design. The full configuration and sources of heat would need to be modeled before the authors would have sufficient confidence to recommend any particular design. In addition, the time taken to cool to base temperature may be an important system parameter. This thermal model only considers the steady state case; a more complete model would be needed to determine the cooldown time, which depends on the mass, heat capacity, and thermal conductances of each item attached to a stage.

Options to improve the thermal performance include replacing the SC-086/50-SCN-CN cable with a microwave flex cable that reduces the static load by scaling the cross-sectional area and combining several signals into one assembly, or with cables made from other low thermal conductivity metals such as brass or stainless steel. In this case, there is a trade-off between reducing the static load due to lower thermal conductivity and increasing the active load due to higher electrical resistivity. Using a superconducting cable would dramatically reduce the static and active heat loads, particularly between the 4K and the MXC stages. This option is possible at the lower fridge stages below the 4K stage, where the temperatures are below the superconducting transition temperatures of commercially available superconducting cables \cite{zotero-367}. The TWPA pump termination resistor could also be relocated to the cold plate, which has a lower heat load. New designs of kinetic inductance parametric amplifiers operate at 4 K and may be able to replace the current LNA and the TWPA \cite{malnou2021}. Optical transduction of the microwave output signal at the mixing chamber \cite{vanthiel2025} is another method that would allow the LNAs to be removed and replace the microwave readout chain with an optical one. Finally, control lines using optical fibers and transducers to create RF signals at the base of the fridge have also been demonstrated and could potentially replace the coax cables \cite{lecocq2021}.

In addition, the required volume, mass, and mounting infrastructure for typical off-the-shelf readout components such as circulators and TWPAs could potentially put an upper bound on the number of readout chains that could fit in the MXC, which in turn would put an upper bound on the number of qubits possible for such a system. Based on experience with the current hardware size, it would be challenging to accommodate more than the 24 readout lines required for the 144-qubit processor on the mixing chamber of a Bluefors XLD1000-SL.

\section{Conclusion}
We measured the thermal conductivity and electrical resistivity of SC-086/50-SCN-CN coaxial cables from room temperature to approximately 4 K. We then used this data to create smooth functions of each respective property with temperature, using simple approximations to extend the fits beyond the experimental data down to the millikelvin regime. We then used these functions to calculate static and active heat loads for a superconducting quantum processor in a dilution fridge in realistic configurations. Finally, we applied these thermal models to a set of hypothetical model flux-tunable superconducting transmon processors of increasing size. The model processors use the tunable coupler architecture, and so each transmon requires two signal lines (an XY control and a flux bias line), while the tunable couplers require a flux bias line. We also modeled the readout system, including TWPAs at the mixing chamber and semiconductor LNAs at the 4K stage. For a Bluefors XLD1000-SL dilution refrigerator using HDW wiring, the theoretical limit of the system using all the available cooling power is approximately 200 qubits. However, with engineering margin in the cooling power and the available space for microwave readout circuitry at the mixing chamber, the practical limit is approximately 140 qubits.

\section{List of Abbreviations}
TWPA, Traveling Wave Parametric Amplifier

LNA, Low Noise Amplifier

MXC, Mixing Chamber

CP, Cold Plate

HDW, Bluefors High-Density Wiring

PTFE, Polytetrafluoroethylene, also known as Teflon

EDX, Energy Dispersive X-ray analysis

\section{Competing Interests}
The authors declare that they have no competing interests.

\section{Funding}
This work was supported by the U.S. Department of Energy, Office of Science, National Quantum Information Science Research Centers, Superconducting Quantum Materials and Systems Center (SQMS) under contract No. DE-AC02-07CH11359..

\section{Availability of Data and Materials}
The thermal conductivity and electrical resistivity datasets are available at https://doi.org/10.5281/zenodo.15238913

\section{Authors' Contributions}
NR, XW, and MF developed the thermal simulation code based on much earlier versions written by those in the Acknowledgments section. MF analyzed the cross-sectional imaging and EDX results. TH measured and analyzed the thermal conductivity. DS and MV measured the electrical resistivity. NR fitted the thermal conductivity and electrical resistivity data. DS, MV, and MH provided fridge data and input on the model quantum processor specifications. NR and MF ran the thermal simulation code. NR, DS, MV, and MF analyzed and interpreted the data. NR, DS, MV, and MF wrote the manuscript. All authors commented on the manuscript. All authors have read and approved the final manuscript.

\section{Acknowledgements}
We would like to thank Jen-Hao Yeh, Saniya Deshpande, and Michael Selvanayagam for their guidance in the early stages of this project. We would also like to thank David Gunnarsson at Bluefors for helpful comments and suggestions on the published performance data of Bluefors dilution refrigerators. 

\clearpage
\raggedright
\bibliography{main}
\end{document}